\documentclass[epj]{svjour}
\usepackage[dvips]{graphicx}
\usepackage{pslatex}
\usepackage{textcomp,amsmath}

\graphicspath{{./}}

\begin{document}

\title{Hartree-Fock-Bogoliubov Calculations of the Rotational Band\\
  of the Very Heavy $^{254}$No Nucleus}
\author{H. Laftchiev\inst{1,2}\and D. Sams\oe{}n\inst{1}\and 
  P. Quentin\inst{1} \and J. Piperova$^{\text{\sf \dag}}$\inst{2}}
\institute{Centre d'\'{E}tudes Nucl\'{e}aires de Bordeaux Gradignan, 
CNRS-IN2P3 and Universit\'{e} Bordeaux-I, BP 120, F-33175 Gradignan, France 
\and
Institute of Nuclear Research and Nuclear Energy (Bulgarian Academy
of Sciences), Tzarigradsko Chaussee 72, 1784 Sofia, Bulgaria}
\dedication{This paper is dedicated to the memory of Jordanka Piperova}
\titlerunning{HFB calculations of the rotational band of $^{254}$No}
\date{}

\abstract{
We report on Hartree-Fock-Bogoliubov (HFB) calculations of the ground-state
rotational band of the heavy nucleus $^{254}$No recently observed
experimentally.
The calculated quadrupole deformation is consistent with the experimental
value of $\beta=0.27$ and is almost constant over the whole band.
We also reproduce fairly well the excitation spectra and moments of inertia of
this isotope up to the maximal experimentally observed state of spin 20.
The rather high stability of this nucleus against fission is illustrated by
the deformation energy curve providing very high fission barriers at
zero spin within the HFB and HFB plus Lipkin-Nogami formalisms. The variation
of these barriers with increased angular velocities is also studied.
}
\maketitle

After the Cohen, Plasil and Swiatecki \cite{CPS} seminal study of the
stability of rotating liquid drops, very heavy nuclei with mass greater than
250 are generally expected to survive rotational excitation only up to rather 
low spins.
Of course, large shell-correction energies added to the bulk liquid drop
estimates may lead to sizable enhancements of the stability against fission in
some cases. Observation of relatively high-spin states in very heavy nuclei
thus provides a very useful tool to assess any theoretical description of
shell effects in this mass region, which is of primary importance for a good
prediction of shell-stabilized superheavy elements.

The experimental observation in the ground-state band of the $^{254}$No
nucleus of rotational states from spins 4 to 18 $\hbar$ (and tentatively 20
$\hbar$) has been reported recently in a series of papers
\cite{Expe:I,Expe:II,Expe:III}. These experiments have been performed using
$^{48}$Ca beams from 130 to 219 MeV on $^{208}$Pb targets. Two of them
(including the more recent) were conducted at the Argonne National Laboratory
with the Gammasphere $4\pi$ Ge detector array, while the other made use of
four Ge clover detectors and was performed at the University of
Jyv\"{a}skyl\"{a}. In all three experiments, the $^{254}$No nuclei were
implanted into a position-sensitive Si strip-detector placed at the focal
plane. The $\gamma$-rays coming from the $^{254}$No were unambiguously
identified from the coincidence with the $\alpha$-decay chain in the Si
detector. As the transition to the ground-state have not been observed, the
spin assignment relies on the use of a Harris parameterization for both
$J^{(1)}$ and $J^{(2)}$ moments of inertia.

In ref. \cite{Expe:III}, the authors proposed a new method to deduce the
fission barrier height $B_f$ based on a reconstruction of the entry
distribution of the evaporation residues in spin and excitation energy. Since
the fission process above the saddle-point energy should be much favored
compared to the $\gamma$-emission, they claim that for any given observed
spin, the maximal excitation energy of the entry distribution should lie below
the saddle-point energy. For each spin the end-point of the excitation energy
distribution gives thus a lower bound for the fission barrier height which is
estimated to be greater than 5 MeV around $I=12$~$\hbar$.

In a recent paper involving some of the authors \cite{Ldm:rot}, it has been
shown using state-of-the-art semiclassical calculations, namely in the extended
Thomas-Fermi (ETF) approach, that the rotating liquid drop model had to be
refined by including a spin-dependence of its parameters which, as a result,
provides a fission stability of heavy nuclei upon increasing the spin which is 
enhanced with respect to what is stated in ref.~\cite{CPS}.
It is the aim of the present paper to go beyond this work, describing
altogether bulk properties as well as shell and pairing effects within the
fully self-consistent microscopic Hartree-Fock-Bogoliubov (HFB) formalism, in
the particular case of $^{254}$No.

In our calculations, we have used the triaxial Hartree-Fock-Bogoliubov code
developed by Laftchiev {\it et al.} \cite{HFB:code,Hristo:these}. This code is
an extension of the Hartree-Fock Routhian code presented in
refs.~\cite{HF:code,Period} which uses Skyrme-type effective interactions and
assumes parity and signature as symmetry operators for the one-body
Hamiltonian.  There, solutions breaking time-reversal symmetry are described
through a decomposition of the single-particle wavefunctions on an axially
symmetric harmonic oscillator basis.  The triaxial character inherent to the
solutions of this formalism is taken care of by a decomposition of the various
densities as Fourier series in the azimuthal angle. This particular choice
provides shorter computation-times than usual triaxial codes as for instance
those of refs.~\cite{Dudu:I,Dudu:II}.
The pairing correlations are described \`{a} la HFB with particle-number
projection tentatively taken into account within the approximate Lipkin-Nogami
(LN) scheme~\cite{Nogami,Kamlah}. The latter method which is widely used for
instance in the context of fully self-consistent microscopic calculations (see
e.g., ref.~\cite{BCSLN} for time-reversal symmetric Hamiltonians, or
ref.~\cite{Gall} for Routhian calculations) results in the addition of a
constraint on the second-order fluctuations of the particle-number. However it
has been shown \cite{Peru} in the $A\simeq190$ mass region from calculations
using the Gogny force that this approximation could produce somewhat 
inconsistent results.

The HFB equations may be cast into a usual eigenvalue form in a doubled space
where the $2\times2$ ``Hamiltonian'' is defined in terms of the Hartree-Fock
Hamiltonian $h$ and of the pairing potential $\Delta$, whereas the
eigensolutions are the usual $U$ and $V$ matrices (see for the notations, e.g,
\cite{RingSchuck}).
Similarly to what was done in ref.~\cite{Gall}, these equations are solved in
two steps at each iteration. First we solve simple Hartree-Fock equations
(that is the HFB equations with vanishing pairing field) to obtain the
eigenstates of the Hartree-Fock--like Hamiltonian. Then the original equations
are solved in a truncated configuration space (through an energy cutoff) of
the Hamiltonian eigenstates previously determined.

For the particle-hole channel, we have used the SkM$^*$ parameterization of
the Skyrme interaction \cite{Skmstar} since it was originally fitted to
provide a good description of the $^{240}$Pu fission barrier and is thus
rather well adapted to describe the stability against fission in very heavy
nuclei. In the particle-particle channel, we have simply used the ``modified 
seniority pairing force'' previously described by Gall \textit{et al.} in
ref.~\cite{Gall}.
At the present stage of our investigations of rotational properties in very
heavy nuclei, we have not attempted to make a global fit over a variety of
different nuclei to provide a pairing force which could be deemed to be
somewhat universal in this mass region. We rather satisfied ourselves with a
parameterization yielding the right moment of inertia at very low spins to
assess the validity of our approach in reproducing the transition energies
(and moment of inertia at higher spins).
The matrix elements $G_n$ and $G_p$ for neutrons and protons are defined with
the usual prescription $G_q=g_q/(11+N_q)$, $N_q$ being the number of particles
in the charge state $q$. The value of $g_n$ ($g_p$ resp.) is 14.3 MeV (15.5
MeV resp.) in the pure HFB case and 10.65 MeV (14.1 MeV resp.) in the HFB+LN
case with an energy cutoff 6.2 MeV above the Fermi level in both cases.  No
smeared boundary conditions for this configuration space have been taken into
account yet which may locally create some minor technical deficiencies.

The calculations have been performed using the usual de\-formation-dependent
truncation scheme equivalent to a 13 major shells spherical basis,
corresponding to more than 450 harmonic oscillator states (for both
signatures).
The basis deformation parameters that have been used in our HFB calculations
were obtained through a minimization of the energy calculated within an
axially and time-reversal symmetric HF+BCS formalism.
At zero angular velocity non axially-symmetrical solutions have been
forcefully searched out (due to the symmetry properties of the interaction,
one would never explore as well known non-axial solutions starting from a
first iteration axially-symmetrical ansatz). At finite spins, the solutions
explicitly break the axial symmetry. As a result however, the solutions are
almost axially symmetric since the $\gamma$ angle never exceeds 1$^\circ$
along the rotational band which allows us to only take into account the 5
lowest-order Fourier series in the densities decomposition (see
ref.~\cite{HF:code}).

Recently, T. Duguet \textit{et al.} \cite{Nobel:bonche} have presented the
results of HFB+LN calculations which could be considered to be in essence
rather similar to ours with two practical differences which are however not
very substantial provided that the technical work is performed adequately [(i)
solutions of the HFB equations are computed on a spatial grid, (ii) a
zero-range force is used in the particle-particle channel]. However they used
for the particle-hole channel the SLy4 parameterization \cite{Sly4a,Sly4b}
which is known to give in average (\textit{i.e.}, semiclassically)
significantly too high fission barriers
\cite{PrivCom}. This may be considered as a drawback to study fission
stability properties.
Another series of calculations for the same nucleus has been published quite
recently \cite{Egido} making use of the D1S Gogny force parameterization
\cite{Gogny:force} which is known to be well suited to study fission barriers.
No corrections for the particle number symmetry violation, e.g., of the
Lipkin-Nogami type, have been taken into account.

Along the rotational band of $^{254}$No, the mass quadrupole moment $Q_0$
varies very slightly between 32.9 and 32.7~b in our calculations. This
corresponds to a deformation parameter of $\beta=0.27$ identical to the
experimental value. It is to be noted that we are there in perfect agreement
with the experimental results \cite{Expe:I,Expe:II,Expe:III} as well as with
the theoretical results of T. Duguet \textit{et al.}  \cite{Nobel:bonche}
yielding $Q_{20}=32.8$~barn.
The calculations of Egido and Robledo on the other hand yield slightly higher
value of $Q_{20}$, resulting in a $\beta$ value close to 0.29.

\begin{figure}[hbt]
  \begin{center}
    \includegraphics[width=0.8400\linewidth]{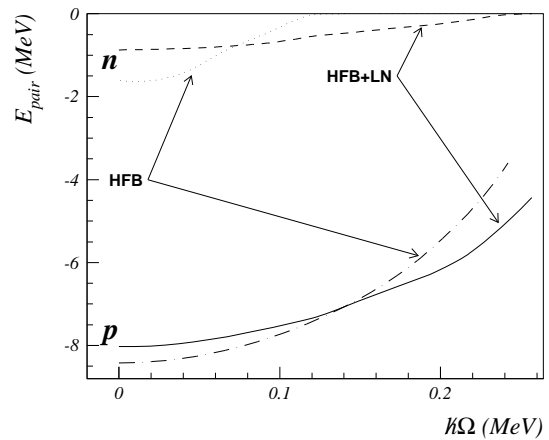}
    \caption{Pairing energies (as defined in text) are plotted as functions
      of the angular velocity. Full and dashed lines (dash-dotted and dotted
      resp.) correspond to proton and neutron energies calculated within the
      HFB+LN (pure HFB resp.) formalism.}
    \label{fig:epair}
  \end{center}
\end{figure}

We present in fig.~\ref{fig:epair} the evolution of the pairing energies,
defined as a quantity proportional to the trace of the product of the abnormal
density $\kappa$ with the pairing potential $\Delta$, for protons and neutrons
with respect to the angular velocity $\Omega$. It is well known from
Nilsson-type calculations that for a deformation parameter of $\beta\simeq0.3$
the $N=152$ neutron number is magic \cite{Magic}. This prediction is
substantiated in our case by the rather low value of the neutron pairing
energy compared to the proton corresponding energy. As usual in HFB
calculations, the pairing energy is seen to decrease with increasing angular
velocity. In particular, the neutron pairing energy vanishes above
$\hbar\Omega= 0.22$ MeV. Within the same angular velocity range, the proton
pairing energy is reduced by more than 40 percent.

It is clear from fig.~\ref{fig:epair}, that around and above rotational
frequencies of the order of $\;\;\hbar\Omega=0.1$~MeV the results of pure HFB
calculations are rather dubious as far as the neutron pairing correlations are
concerned. This remark is clearly applicable also to the calculations of Egido
and Robledo.  It is generally expected that the Lipkin-Nogami corrected HFB
calculations should be better adapted to such a situation of weak pairing
correlations. However, as already mentioned, the calculations of Peru {\it et
al.}  (See, e.g., \cite{Peru}) in the superdeformed $A\simeq190$ region show
that the quality of the results of such an approach is rather unpredictable.
The application to such a Routhian approach of pairing calculations conserving
explicitly the particle number, e.g., as those of Pillet {\it et al.}
\cite{Pillet,Pillet:phd} would be of great interest there. In view of this
expected inadequacy of pure HFB approach we have therefore limited the
discussion below of calculated moments of inertia to HFB+LN results.

\begin{figure}[hbt]
  \begin{center}
    \includegraphics[width=0.8400\linewidth]{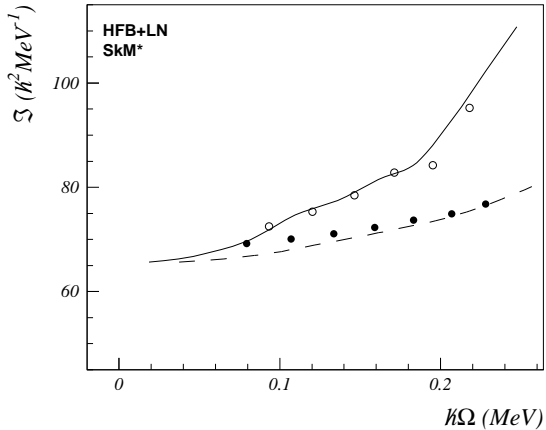}
    \caption{Moments of inertia (see text) are plotted as functions of the 
      angular velocity. Dynamic (kinematic resp.) moment of inertia calculated
      within the HFB+LN formalism is displayed in full line (dashed resp.)
      while its experimental counterpart is represented by open circles (full
      resp.).}
    \label{fig:inertia}
  \end{center}
\end{figure}

The kinematic ($J^{(1)}$) and dynamic ($J^{(2)}$) moments of inertia are
obtained from our calculations using the formulae
\begin{eqnarray}
  \label{eq:inertia}
  J^{(1)} &= \frac{\langle\hat j_x \rangle}{\Omega}\\
  J^{(2)} &= \frac{\partial \langle\hat j_x \rangle}{\partial \Omega}
  \label{moment-j2}
\end{eqnarray}
where $\hat j_x$ is the total angular momentum operator and $\Omega$ is the
angular velocity. The derivative involved in the $J^{(2)}$ calculations have
been taken over two consecutive even-spins.
Both are plotted in fig.~\ref{fig:inertia} as functions of
the angular velocity together with their experimental counterparts.

The kinematic moment of inertia $J^{(1)}$ is, in fact, the expression of the
linear response of the system under the constraint of some collective
variable. Nevertheless, we have actually assumed that it could be computed
from the wavefunctions yielding the corrected energy (i.e., bypassing an
explicit treatment of LN corrections specific to $\langle j_x\rangle$ in the
spirit of ref.  \cite{BCSLN}).
It may also be argued that due to the dependence of $\lambda_2$ upon the
angular velocity $\omega$, equation (\ref{moment-j2}) is not correct and
dynamic moments of inertia should be evaluated through second derivatives of
the energy with respect to the spin. However, we have checked that this
$\omega$-dependence is very small and can be safely neglected (at least up to
about $\omega=0.1$ MeV).

\begin{figure}[h]
  \begin{center}
    \includegraphics[width=0.8400\linewidth]{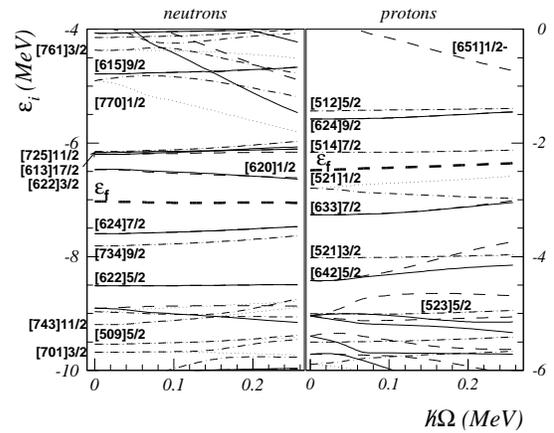}
    \caption{Single-particle Routhians in the $^{254}$No nuclei for neutrons
      (left panel) and protons (right panel) as a function of the angular
      velocity. The convention used for the (parity, signature) representation 
      is the following: (+,+) in full lines, (+,-) in dashed lines, (-,+) in
      dash-dotted lines and (-,-) in dotted lines. Calculations are performed
      within the HFB+LN formalism.}
    \label{fig:spe}
  \end{center}
\end{figure}

It is seen on fig. \ref{fig:inertia} that we are reproducing rather well the
experimental trend for both the $J^{(1)}$ and $J^{(2)}$ moments of
inertia. The agreement of our results is better than the one obtained by Egido
and Robledo for the only moment which they report, namely $J^{(1)}$. It is
also better than the only moment reported by Duguet {\it et al.}, namely
$J^{(2)}$ in this case. The latter is quite remarkable because, as it can be
seen on fig. \ref{fig:spe}, our single particle spectra are somewhat different
for the neutron states. Whereas these authors find important deformed gaps
feature around $N=150$ and $N=152$ (see fig. 5 in \cite{Nobel:bonche}), we get
only a comparable $N=152$ gap plus a $N=170$ gap at low spins which is not
present in the paper of Duguet {\it et al.}.

\begin{figure}[htb]
  \begin{center}
    \includegraphics[width=0.8400\linewidth]{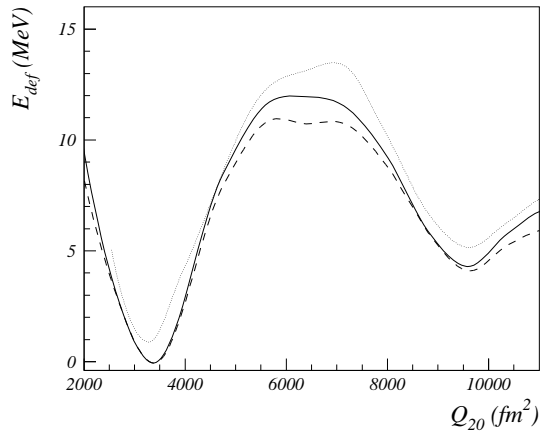}
    \caption{Deformation energy curves are plotted as functions of the
      quadrupole moment $Q_{20}$. Dashed line corresponds to HFB calculation
      at zero-spin, while full line (dotted resp.) corresponds to HFB+LN
      calculations at zero-spin (12~$\hbar$ resp.). The energy reference is
      taken as the unconstrained zero-spin energy value in both formalisms.}
    \label{fig:barrier}
  \end{center}
\end{figure}

On fig. \ref{fig:barrier} we display the fission barriers obtained at zero
total angular momentum within the HFB and HFB+LN approaches allowing
triaxial deformation. 
The former yields a first fission barrier which is slightly lower than what is
obtained for the latter. Comparing HFB+LN results now for two different total
angular momenta $I=0$~$\hbar$ and $I=12$~$\hbar$ we find that the first
fission barrier height is slightly increased by 500 keV where one would expect
it to be decreased.
This phenomenon is due to a higher single-particle level density at the top of
the first barrier for $I=12$~$\hbar$ as compared to what is obtained at
zero-spin. At such low spins, shell structure effects are thus able to mask
the anti-binding effect of the rotation leading to an extra fission
instability
With the basis size in use in our calculations we are not a
priori able to provide reliable relative energies in the second fission
barrier region, as noted years ago \cite{Flocard:basis}.

To conclude it appears that the SkM$^*$ parameterization of the Skyrme
interaction (complemented by a seniority force in the p-p h-h channel) is able
to provide a good reproduction of the spectroscopic data in the ground
state rotational band of the $^{254}$No nucleus within the HFB plus
Lipkin-Nogami approach.
This result is all the more remarkable that similar calculations
\cite{Nobel:bonche} using a different effective force and yielding a rather
different shell structure in the relevant deformation range, lead to a
similarly (may be slightly less) good reproduction of the experimental data.
Even though it is true that reproducing such collective rotation data provides
a much wanted test of effective forces to be used when predicting the
stability properties of neighbouring superheavy nuclei, such a convergence of
results from seemingly rather different effective forces should mitigate the
hope to get so far here a completely stringent benchmark.

The calculations of ref. \cite{Egido} have studied in detail the fission
properties of this nucleus, namely they underline its remarkable, and somewhat
unexpected, fission stability which could partly be explained by a change of
saturation properties of nuclear matter in presence of a centrifugal field as
advocated in ref. \cite{Ldm:rot}. Our results, limited in both spins and
deformations as compared to those of ref. \cite{Egido} yield similar
conclusions on this stability.
A pending point common to all these calculations (those of refs.
\cite{Nobel:bonche,Egido} as well as ours) is the pairing correlation treatment
in weak pairing regions as encountered upon increasing the angular
velocity. Surely, treatment of such correlations within an approach that
explicitly conserves the particle number, as the one proposed in
refs.\cite{Pillet,Pillet:phd}, would greatly improve the validity of the
results.

However whatever their mid stream character, the rather satisfactory
reproduction of the difficult, and for that reason somewhat scarce,
experimental results so far obtained by self consistent calculations as those
of refs. \cite{Nobel:bonche,Egido} and the one presented here, should
constitute an incentive to experimentalists to provide more data so as to
improve our knowledge of the effective interaction to be used. This certainly
constitutes a very timely endeavour.


\vfill
\end{document}